\documentclass[letterpaper]{article} % DO NOT CHANGE THIS
\usepackage{aaai25}  % DO NOT CHANGE THIS
\usepackage{times}  % DO NOT CHANGE THIS
\usepackage{helvet}  % DO NOT CHANGE THIS
\usepackage{courier}  % DO NOT CHANGE THIS
\usepackage[hyphens]{url}  % DO NOT CHANGE THIS
\usepackage{graphicx} % DO NOT CHANGE THIS
\usepackage{natbib}  % DO NOT CHANGE THIS AND DO NOT ADD ANY OPTIONS TO IT
\usepackage{url}
\usepackage{xcolor}
\usepackage{amssymb}

\usepackage{caption} % DO NOT CHANGE THIS AND DO NOT ADD ANY OPTIONS TO IT
\usepackage{algorithm}
\usepackage{algorithmic}
\usepackage{float}
\usepackage{xcolor}
\usepackage{amsmath}
\newcommand{\ye}[1]{{\color{black}#1}}

\usepackage{newfloat}
\usepackage{listings}
\DeclareCaptionStyle{ruled}{labelfont=normalfont,labelsep=colon,strut=off} % DO NOT CHANGE THIS
\floatstyle{ruled}
\newfloat{listing}{tb}{lst}{}
\floatname{listing}{Listing}
\title{SigStyle: Signature Style Transfer via Personalized Text-to-Image Models} 
\author{
    Ye Wang\textsuperscript{\rm 1},
    Tongyuan Bai\textsuperscript{\rm 1},
    Xuping Xie\textsuperscript{\rm 2},
    Zili Yi\textsuperscript{\rm 3},
    Yilin Wang\textsuperscript{\rm 4}\footnote{Corresponding authors},
    Rui Ma\textsuperscript{\rm 1,5}\footnotemark[1]
    }
\affiliations{
    \textsuperscript{\rm 1} School of Artificial Intelligence, Jilin University\\
    \textsuperscript{\rm 2} College of Computer Science and Technology, Jilin University\\ 
    \textsuperscript{\rm 3} School of Intelligence Science and Technology, Nanjing University,
    \textsuperscript{\rm 4} Adobe\\
    \textsuperscript{\rm 5}  Engineering Research Center of Knowledge-Driven Human-Machine Intelligence, MOE, China\\
    \{yewang22, baity23, xiexp21\}@mails.jlu.edu.cn, yi@nju.edu.cn, yilwang@adobe.com, ruim@jlu.edu.cn
    \\
    Project Page:
    \url{https://wangyephd.github.io/projects/sigstyle.html}
}

\usepackage{bibentry}
\usepackage{cuted} % 支持跨栏内容
\begin{document}

\maketitle

\begin{figure*}[t]
    \centering
    \includegraphics[width=0.9\textwidth]{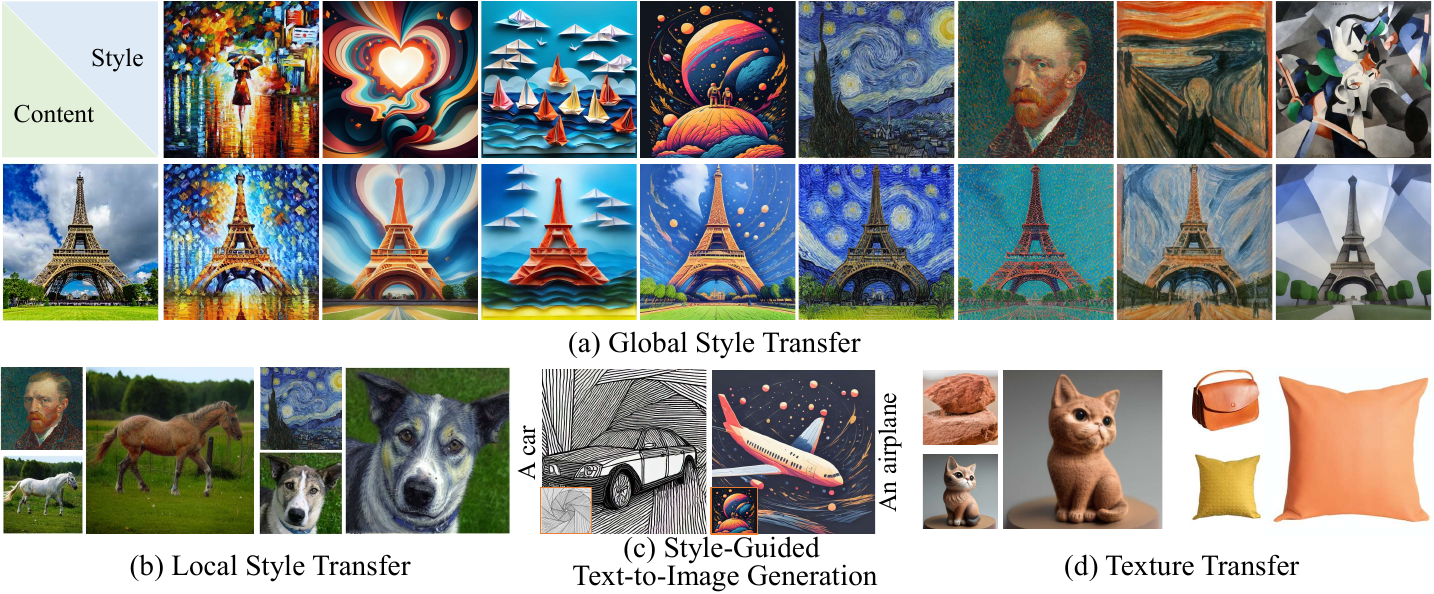}
    \caption{Our method can achieve high-quality global style transfer (a) while keeping the signature style such as distinct and recognizable visual traits like geometric and structural patterns, color palettes and brush strokes etc. Also, our method is flexible and supports local style transfer (b), style-guided text-to-image generation (c), and texture transfer (d). Best viewed \\ in color.}
    \label{teaser}
\end{figure*}
% \section{Method}

\begin{abstract}
Style transfer enables the seamless integration of artistic styles from a style image into a content image, resulting in visually striking and aesthetically enriched outputs. 
Despite numerous advances in this field, existing methods did not explicitly focus on the \textit{signature style}, which represents the distinct and recognizable visual traits of the image such as geometric and structural patterns, color palettes and brush strokes etc.
In this paper, we introduce SigStyle, a framework that leverages the semantic priors that embedded in a personalized text-to-image diffusion model to capture the signature style representation. This style capture process is powered by a hypernetwork that efficiently fine-tunes the diffusion model for any given single style image. Style transfer then is conceptualized as the reconstruction process of content image through learned style tokens from the personalized diffusion model. Additionally, to ensure the content consistency throughout the style transfer process, we introduce a time-aware attention swapping technique that incorporates content information from the original image into the early denoising steps of target image generation. Beyond enabling high-quality signature style transfer across a wide range of styles, SigStyle supports multiple interesting applications, such as local style transfer, texture transfer, style fusion and style-guided text-to-image generation. Quantitative and qualitative evaluations demonstrate our approach outperforms existing style transfer methods for recognizing and transferring the signature styles. 
\end{abstract}


\section{Introduction}

% \ye{The following setcion }
Style transfer technology \cite{gatys2016image}, which seamlessly incorporates stylistic elements into content images to produce visually impactful results, has gained widespread attention in recent years due to its extensive applications in art design, photography, fashion and other fields. A critical aspect of style transfer is the preservation of the original style during the transfer process. Ensuring that these intricate details and artistic expressiveness are essential for achieving high-quality results, especially when dealing with complex styles such as the  shape and layout of artistic elements, and the patterns of brush strokes and lines.

Early style transfer methods primarily include techniques relying on local region matching \cite{zhang2013style,wang2004efficient}, or use convolutional neural network (CNN) \cite{gatys2016image, gatys2017controlling, kolkin2019style} or feed-forward network \cite{deng2020arbitrary,huang2017arbitrary,liao2017visual,zhang2022exact} to achieve style transfer. 
With the rapid advancement of diffusion models, diffusion-based style transfer has significantly progressed. These methods can be categorized into two types: tuning-based and tuning-free. A representative tuning-based method is InST \cite{zhang2023inversion}, which introduces a textual inversion-based approach that maps a given single reference style image into a corresponding textual embedding. This textual embedding is then used as a condition to achieve style transfer for the content image. 
In contrast, tuning-free methods such as StyleID \cite{chung2024style}, DiffStyle \cite{jeong2023training}, InstantStyle \cite{wang2024instantstyle}, InstantStyle-Plus \cite{wang2024instantstyleplus}, and FreeTuner \cite{xu2024freetuner} merge style and content features extracted from the attention layers of Stable Diffusion \cite{rombach2022high} to achieve style transfer. These methods offer superior computational efficiency and only require a single forward pass without the need for additional tuning.

\begin{figure}[t]
    \centering
    \includegraphics[width=0.9\linewidth]{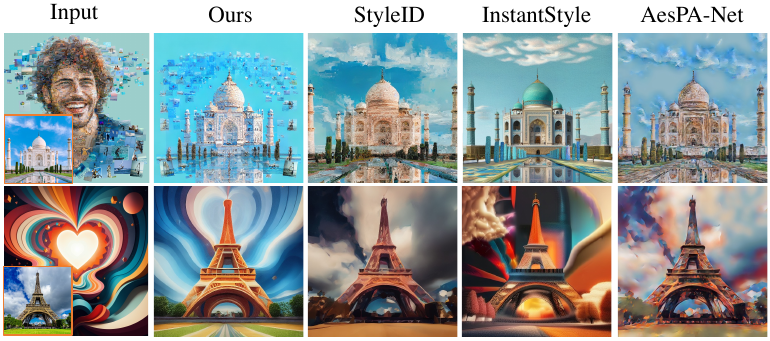}
    \caption{Signature style transfer comparison with SOTA methods on two complex style references.}
    \label{fig:complex_style}
\end{figure}

Despite the significant progress of the aforementioned methods, signature style transfer remains underexplored. Signature style refers to the unique and recognizable visual traits that defines a particular 
artistic style, such as geometric and structural patterns, color palettes, and brush strokes. 
%style, including geometric and structural patterns, color palettes and brush strokes etc. 
For example, as illustrated in the first row of Figure \ref{fig:complex_style}, the signature style of this image is defined by the structural arrangement and composition of numerous small images that together form the figure of a person. Additionally, the signature style of the image in the second row is characterized by geometric and structural patterns, as well as distinctive color palettes.
% The signature style is more challenging to preserve during the transfer process.
Although existing methods often succeed in transferring basic color information, they fail to capture and retain the essential artistic style from the reference images, including small image blocks, colorful ribbon-shaped lines and other intricate characteristics as shown in Figure \ref{fig:complex_style}.
This highlights a critical limitation: current methods struggle to achieve signature style transfer.

% 

% they still fall short in perfectly preserving complex styles, \ye{including lines, color textures, layouts, semantics, structures, and object elements}.

The main reason for the aforementioned issues is that existing methods insufficiently consider the distinct visual traits of style images.
Meanwhile, 
personalized text-to-image models such as Dreambooth \cite{ruiz2023dreambooth} can capture rich information about style concepts including structures, pose, texture, lines, and more. Inspired by this, we propose SigStyle, a novel framework that leverages rich priors embedded in a personalized text-to-image model to capture complex style for facilitating signature style transfer. However, existing customized text-to-image models often require multiple reference images for fine-tuning, making them unsuitable for style transfer tasks that depend on a single reference image. To address this limitation, we propose a hypernetwork-powered, style-aware fine-tuning mechanism that enables precise concept learning and accurate inversion of style attributes using just one style reference image.

%On the other hand, existing customized text-to-image models typically require multiple reference images for fine-tuning. It is not suitable for style transfer tasks that rely on a single reference image. Therefore, we introduce a hypernetwork-powered style-aware fine-tuning mechanism that enables precise concept learning and the inversion of style attributes given a single style reference image.

Specifically, we employ a lightweight hypernetwork to modulate and refine diffusion UNet weights. This strategy can not only ensure smoother updates of the parameters and reduces the likelihood of overfitting, but also effectively capture and represent the signature style attributes from a single reference image. Furthermore, unlike existing fine-tuning approaches \cite{zhang2023inversion, ruiz2023dreambooth, zhang2023inversion}, our method focuses on fine-tuning only the modules related to style attributes, instead of the entire diffusion network. Such fine-tuning mechanism not only enhances tuning efficiency, but also improves the style inversion accuracy. Based on this fine-tuning mechanism, we can represent a signature style as a special token *.
Subsequently, we define style transfer as the reconstruction process of the content image based on the style token learned from the customized diffusion model. Additionally, to maintain content consistency, we propose a time-aware attention swapping technique inspired by PhotoSwap \cite{gu2024photoswap} and SwapAnything \cite{gu2024swapanything}. This technique transfers content-related attention from the original image generation process to the target image generation during the early denoising timesteps, ensuring content consistency throughout the style transfer process.

As shown in Figure \ref{teaser}, our method achieves high-quality signature style transfer results across various complex style references. Moreover, our framework not only supports global style transfer, but also supports a broad range of applications, including local style transfer, texture transfer, style fusion, and style-guided text-to-image generation. Extensive experiments and evaluations further demonstrate the versatility and effectiveness of our method.

Our contributions can be summarized as follows:

\begin{itemize}
    \item We propose SigStyle, a novel framework that is the first to explicitly focusing on the challenging signature style transfer via utilizing the personalized text-to-image diffusion models.
    
    % . \yilin{most style transfer methods use single style image. We should emphasize we are the first work to consider the style preservation.}
    \item  We introduce a hypernetwork-powered style-aware fine-tuning mechanism that can learn the signature style attributes from only a single style image. This approach overcomes the limitation of customization methods that need multiple reference images, making it more suitable for style transfer tasks.
    % that not only eliminates severe overfitting issues associated with single-image fine-tuning but also enables precise learning and inversion of style attributes. Combined with a time-aware attention swapping technique, this allows efficient style transfer while maintaining content consistency.  \yilin{existing personalize text to image model can capture the richful information from the referece concepts including pose texture ans structures. But they often require a few images for finetuing which is not suitable for style transfer. Thus we propsoed hyper network.}
    \item Extensive experiments show that our method outperforms existing state-of-the-art methods on style transfer. Notably, our approach excels in preserving signature style details, highlighting its superior capability in signature style transfer. Moreover, the diverse applications emphasize the generalizability and versatility of our method.
    % \yilin{Demonstrate that we can keep the style very well others can't.}
\end{itemize}

\section{Related Work}

\paragraph{\textbf{Style Transfer.}}
Style transfer \cite{zhang2019multimodal,wang2020diversified,wang2020collaborative,park2019arbitrary,lu2019closed,li2018closed,li2017universal,lai2017deep,gatys2016image} applies the artistic style of one image to another while preserving the latter's content and structure. Early methods \cite{zhang2013style,wang2004efficient} relied on handcrafted features, while CNN-based approaches \cite{gatys2016image, gatys2017controlling, kolkin2019style} leveraged pre-trained networks to capture style pattern. Arbitrary style transfer methods \cite{deng2020arbitrary,huang2017arbitrary,liao2017visual,zhang2022exact} further improved this by using unified feed-forward models for flexible inputs.

In recent years, diffusion models have increasingly been employed for style transfer. These methods can be broadly categorized into two types: tuning-based methods and tuning-free methods. A representative example of the former is InST \cite{zhang2023inversion}, which introduces a text inversion-based approach, aiming to map a given style to its corresponding text token. 
% However, this method lacks precision in style inversion, often altering the content of the original image during style transfer. 
StyleDiffusion \cite{wang2023stylediffusion} refines diffusion models by introducing a CLIP-based style decoupling loss, effectively separating style from content. 
% FreeTuner \cite{xu2024freetuner} proposes compositional personalization, allowing the combination of different subject and style concepts. However, this approach often alters the structure and content of the original image.
Tuning-free methods achieve style transfer in a single forward process without model fine-tuning. DEADiff \cite{qi2024deadiff} utilizes the Q-Former \cite{li2023blip} and paired datasets to extract decoupled representations of content and style, facilitating style transfer. InstantStyle \cite{wang2024instantstyle} uses specific block injection techniques to implicitly decouple content and style for effective style transfer. Building on this, InstantStyle-Plus \cite{wang2024instantstyleplus} introduces ControlNet \cite{zhang2023adding} to further maintain the integrity of image content. StyleID \cite{chung2024style} adjusts self-attention layers and introduces novel techniques such as query preservation and initial latent AdaIN to maintain content integrity.

Nevertheless, these approaches struggle with achieving signature style transfer. In this paper, we address this challenge by utilizing a personalized text-to-image model to perform signature style transfer, generating visually compelling and aesthetically enhanced outputs.

\paragraph{\textbf{Personalized Text-to-Image Generation.}} 

Recent studies \cite{ruiz2023dreambooth, gal2022image, ruiz2023hyperdreambooth, kumari2023multi} have increasingly focused on using visual exemplars as a foundation for image generation to address the inherent ambiguity and unpredictability of text-based prompts. This approach involves collecting multiple reference images to fine-tune diffusion models. 
However, for style transfer tasks that rely on a single style image, the aforementioned methods often cause severe overfitting when fine-tuning on this single image, significantly diminishing the effectiveness of the style transfer.
In contrast, our proposed hypernetwork-powered style aware fine-tuning method can overcome the limitations associated with fine-tuning on a single style image. It enables precise inversion of style attributes without the drawbacks of overfitting
% \yilin{will single image based personalized method work?}

% In addition to customizing the general essence of the reference image, some works have begun to explore attribute-level T2I customization \cite{voynov2023p+, zhang2023prospect, goel2023pair, yang2023paint} . However, these methods either require multiple reference images \cite{voynov2023p+} or extensive fine-tuning \cite{goel2023pair,yang2023paint, ye2023ip} across large datasets. 

\paragraph{\textbf{Parameter Efficient Fine Tuning (PEFT).}} PEFT represents an innovative approach in the refinement of deep learning models, emphasizing the adjustment of a subset of parameters rather than the entire model. These parameters are identified as either specific subsets from the originally trained model or a minimal number of newly introduced parameters during the fine-tuning phase. PEFT has been applied in text-to-image diffusion models \cite{saharia2022photorealistic,rombach2022high} through techniques such as LoRA \cite{ryu2023low} and adapter tuning \cite{mou2023t2i,ye2023ip,wei2023elite,chen2024subject,ma2023unified}. In this paper, we leverage a hypernetwork to adjust and refine a unique subset of pre-trained parameters.

\begin{figure*}[t]
    \centering
    \includegraphics[width=0.9\textwidth]{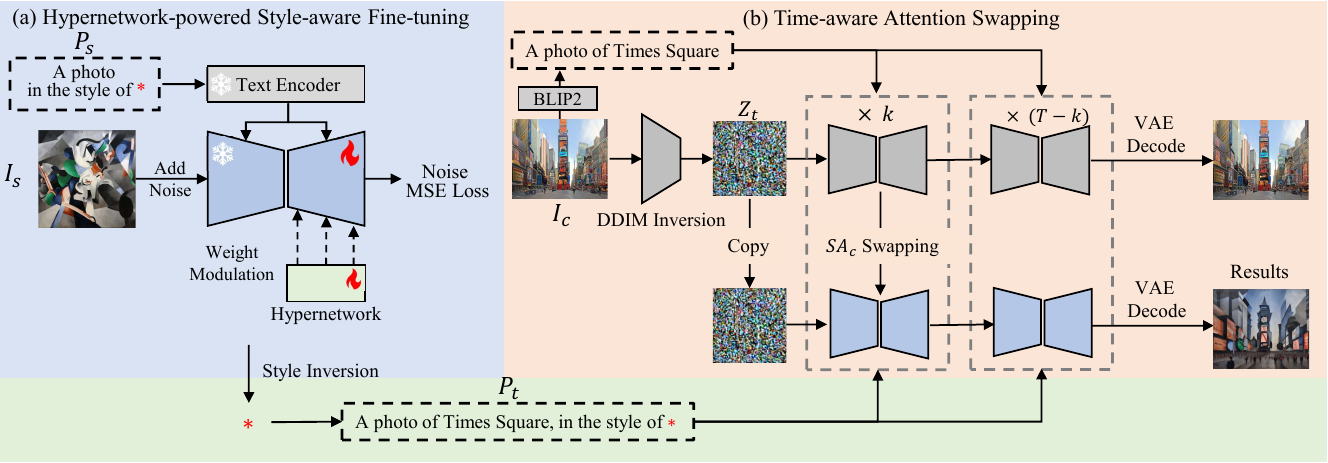}
    \caption{The SigStyle framework. First, given a style image, we perform hypernetwork-powered style-aware fine-tuning for style inversion and represent the reference style as a special token * (see Figure \ref{fig:method}.a). In Figure \ref{fig:method}.b, the upper branch represents the reconstruction process of the content image, while the lower branch represents the generation process of the target image. When generating the target image using a pre-trained model and target text, we first use DDIM Inversion to map the content image into noise latents, which are then copied as the initial noise for generating the target image. Then, we adopt time-aware attention swapping to inject structural and content information during the first $k$ steps of the denoising process (see Figure b). In the subsequent $T-k$ steps, we proceed with the usual denoising process without any swapping. Finally, by decoding with VAE, we obtain the style-transferred image. }
    \label{fig:method}
\end{figure*}
\section{Method}

\subsection{Preliminaries}

\paragraph{\textbf{Stable Diffusion.}} 
Stable Diffusion \cite{rombach2022high} is a state-of-the-art text-to-image model that operates in a low-dimensional latent space. It encodes an image $x$ into a latent $z$ via a VAE encoder, adds noise $\epsilon$ at time step $t$ to obtain a noisy latent $z_t$, and uses a CLIP text encoder $\tau$ to incorporate textual prompts $c$ through cross-attention layers. A conditional UNet $\epsilon_\theta$ is then trained to predict the noise $\epsilon$, guided by the following training objective:
% Stable Diffusion \cite{rombach2022high}, a state-of-the-art text-to-image generation model, operates within a low-dimensional latent space. It begins by encoding an input image $x$ into a latent representation $z$ using a VAE encoder. Noise $\epsilon$ is then introduced at time step $t$ to create a noisy latent $z_t$. To guide the generation process with text conditions, Stable Diffusion incorporates a CLIP text encoder $\tau$ to encode textual prompts $c$, which are integrated into the cross-attention layers for interaction with the noisy latents. Finally, a conditional U-Net backbone $\epsilon_\theta$ is trained to predict the noise $\epsilon$. The training objectives is as follows:
\begin{equation}
    L_{SD}(\theta) := \mathbb{E}_{t,x_0,\epsilon} \left[ \lVert \epsilon - \epsilon_\theta(z_t, t, \tau(c)) \rVert^2 \right] \label{eq:LLDM}.
\end{equation}

% Classifier-free guidance \cite{ho2022classifier} is employed to steer the efficient inference of Stable Diffusion. It is designed to extrapolate the output of the model in the direction of $\epsilon_\theta(x_t \mid \tau(c))$
% and away from $\epsilon_\theta(x_t \mid \oslash)$ as follows:

% \begin{equation}
% \hat{\epsilon}=\epsilon_\theta\left(x_t \mid \oslash\right)+s \cdot\left(\epsilon_\theta\left(x_t \mid \tau(c)\right)-\epsilon_\theta\left(x_t \mid \oslash\right)\right)
% \end{equation}
% $\oslash$ is a null text, s is the guidance weight and increasing $s > 1$ 
% strengthens the effect of guidance.

\paragraph{\textbf{Attention Mechanisms in Diffusion.}}

The UNet-based diffusion model comprises self-attention and cross-attention modules. As demonstrated in \cite{hertz2022prompt, tumanyan2023plug, gu2024photoswap, gu2024swapanything}, self-attention maps capture image structure and identity-related information. The computation of self-attention maps is defined as follows:
\begin{equation}
S A=\operatorname{Softmax}\left(\frac{Q_s K_s^T}{\sqrt{d}}\right),
\end{equation}
where $Q_s$ and $K_s$ represent different projections of visual features, and $d$ is the dimension of feature used for scaling.

% \paragraph{\textbf{DDIM Inversion}}

\subsection{SigStyle}

\paragraph{\textbf{Pipeline Overview.}}
Given a style image $I_{s}$ and a content image $I_{c}$, SigStyle can seamlessly transfer the signature style to the content image while preserving the original content. The SigStyle pipeline is illustrated in Figure \ref{fig:method}. First, we use a hypernetwork to perform style-aware fine-tuning on diffusion model with a single style image and represent the target style using a special token $*$ (see Figure \ref{fig:method}.a).
Next, we employ DDIM Inversion to obtain the noise $z_{t}$ that can reconstruct the content image. Then, we extract the required self-attention maps $SA_{c}$ from the UNet, which preserve the structure and content information of the content image. Finally, during the generation of the target image that conditioned on the noise $z_{t}$ and the target text prompt $P_{t}$, we use the obtained self-attention maps to replace the corresponding ones in the first $k$ steps to maintain the content information (see Figure \ref{fig:method}.b ).

\paragraph{\textbf{Hypernetwork-powered Style-aware Fine-tuning.}} 
Object-level inversion \cite{ruiz2023dreambooth} typically requires fine-tuning the entire UNet, but such an approach is not suitable for style inversion. We hypothesize that style, as an attribute of the image, should be learned and understood by specific modules within the network. To verify this hypothesis, we conducted an in-depth analysis of the style attribute learning preferences of UNet.

\begin{figure}[t]
    \centering
    \includegraphics[width=0.8\linewidth]{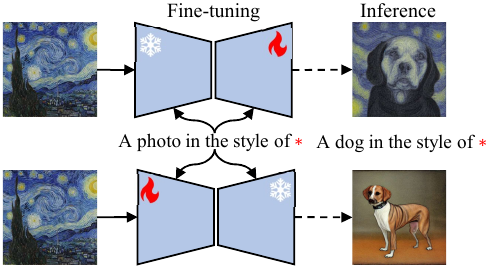}
    \caption{Style learning preferences analysis of UNet's encoder and decoder.}
    \label{fig:style_tuning}
\end{figure}

% \paragraph{\textit{Style Learning Preferences Analysis of UNet.}}
\textit{Style Learning Preferences Analysis of UNet.}
Previous work \cite{voynov2023p+, zhang2023prospect} has also attempted to analyze the learning preferences of different layers, focusing mainly on shape and color. However, these studies did not accurately identify the specific layers responsible for learning style attributes. Therefore, we conducted a simple experiment using Stable Diffusion to analyze the learning preferences of different modules of UNet. As shown in Figure \ref{fig:style_tuning}, we first divided the UNet into two parts: the encoder and the decoder. we selected a style reference image, for example, \textit{The Starry Night}, and inputted default text prompts such as ``a photo in the style of *." The * is learnable token embedding. Subsequently, we separately fine-tuned the encoder and decoder modules of the UNet and utilized the fine-tuned model to generate the corresponding images. \ye{During fine-tuning, * is progressively refined to encapsulate the distinctive style patterns in the style image. This enables * to function as an abstract, high-level representation of the style, effectively acting as a trigger for the fine-tuned T2I model to generate stylized outputs. Importantly, the original style image is not required as input once the model has been fine-tuned, as * captures the essential elements of the style.}
During inference,
The text prompt used for inference is ``a dog in the style of *". 

We observe significant differences in the generated results when separately fine-tuning the encoder and decoder. Fine-tuning the encoder fails to produce images that match the style of the reference image, while fine-tuning the decoder results in images with similar styles. This experiment further validates our hypothesis that style attributes are learned by specific network modules, namely the decoder module.

Based on this insight, we propose a style-aware fine-tuning mechanism that tunes only the decoder module of the UNet to learn style attributes. However, this approach may suffer from severe overfitting issues, particularly when fine-tuning with a single image. To address this challenge, we use a hypernetwork to refine and modulate the network parameters, achieving smoother updates and thereby reducing the risk of overfitting.

\textit{Hypernetwork.}
The hypernetwork architecture is  shown in the Figure \ref{fig:hypernetwork}, we draw inspiration from E4T \cite{gal2023encoder} to design our hypernetwork.
The module takes as input a learnable constant $cons$ (default-initialized to 1) and the dimension information $[dim_{r}, dim_{c}]$ of the target weight parameters. It is then trained to predict weight offsets in the same dimensions as the target weight parameters. Here, $dim_{r}$ represents the number of rows of the target weight parameters, and $dim_{c}$ represents the number of columns.
In detail, the learnable constant passes through two linear layers, yielding outputs that are multiplied to derive the initial weight offset matrix. Row and column transformations are then applied to this matrix to obtain the final weight offset matrix $\Delta w$. As discussed in the literatures \cite{gal2023encoder,kumari2023multi,wei2023elite}, the weights of self-attention and cross-attention play a crucial role in the process of image customization.
Therefore, we utilize the hypernetwork as a weight offsets prediction module to modulate and guide the updates of attention-related weights within the decoder.
The high-level parameter update process is defined as follows:
\begin{figure}[t]
    \centering
    \includegraphics[width=0.8\linewidth]{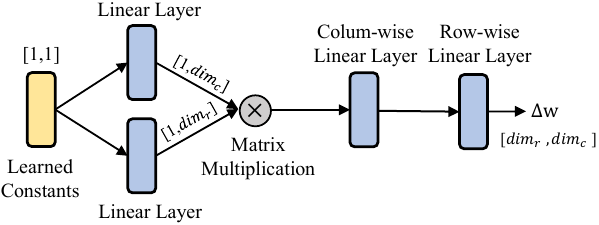}
    \caption{The architecture of hypernetwork.}
    \label{fig:hypernetwork}
\end{figure}
\begin{align}
    \Delta w &= hypernetwork(cons,dim_{r},dim_{c}),  \\
    w^*_{attn} &= w_{attn} + \lambda * \Delta w,
\end{align}
where $w_{attn}$ denotes the general term for attention-related parameters, including the query matrix, key matrix and value matrix for self-attention and cross-attention layers; $\lambda$ is a weight coefficient that is used to regulate the updating strength of parameters. During training, $\lambda$ is set to 1.0. 

\paragraph{Loss Function.} 
To guide the style inversion, we adapt the original noise prediction loss function to work with the text prompt for style learning:
\begin{equation}
    Loss(\theta) := \mathbb{E}_{t,x_0,\epsilon} \left[ \lVert \epsilon - \epsilon_\theta(x_t, t, \tau(P_s)) \rVert^2 \right]. 
    \label{eq:LLDM}
\end{equation}
Note that $\theta$ denotes the parameters of the decoder and hypernetwork, $t$ denotes the current time step, $\epsilon$ denotes the noise, $x_t$ represents the noise latent of style image at time $t$,  $\tau(P_s)$ represents the CLIP embedding of the input text prompt, i.e., ``a photo in the style of *.''

\begin{figure*}[t]
    \centering
    \includegraphics[width=0.85\textwidth]{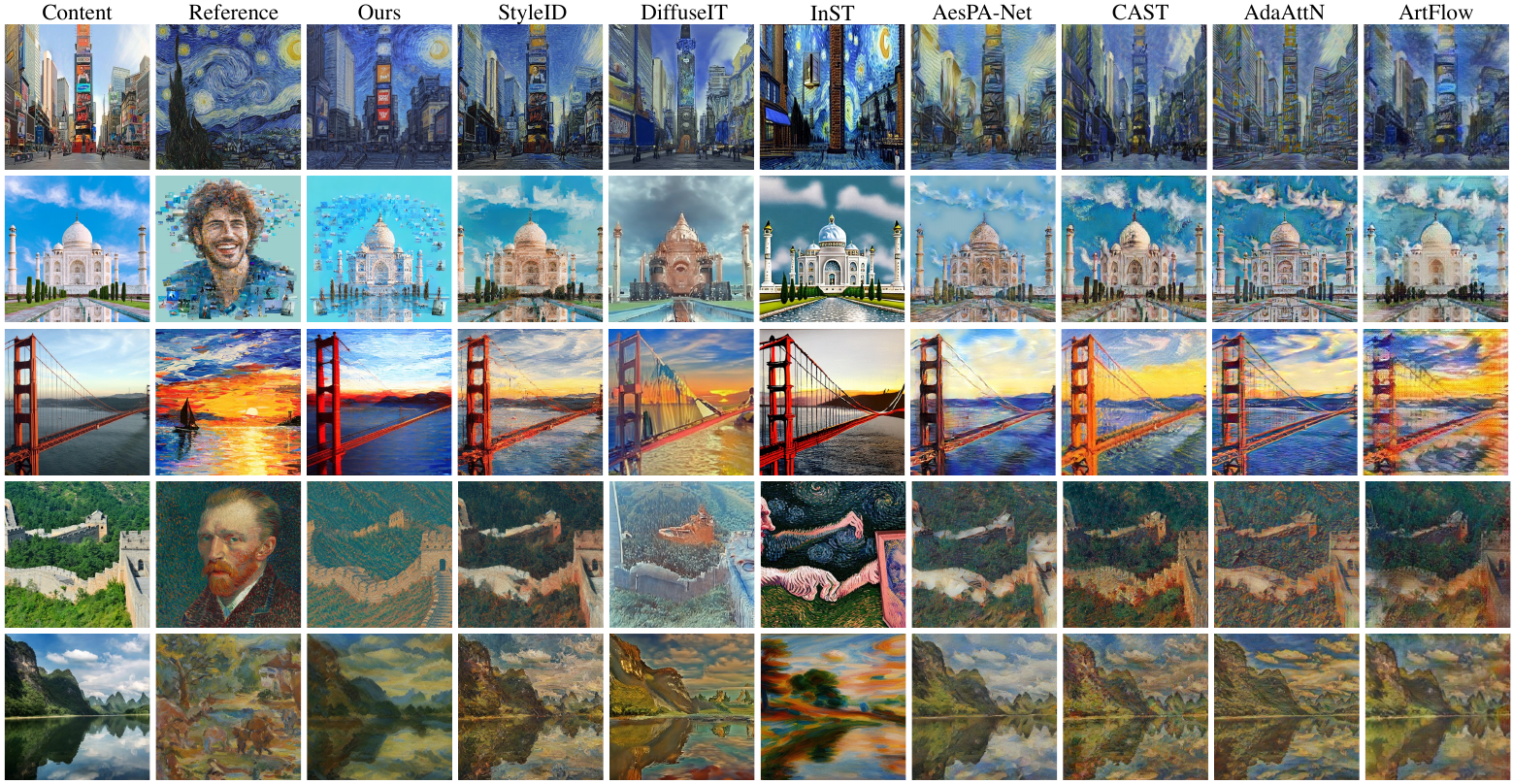}
    \caption{ Qualitative comparison with various SOTA image style transfer methods for global style transfer. }
    \label{fig:compare_other}
\end{figure*}

\paragraph{\textbf{Time-aware Attention Swapping.}}
To achieve style transfer, a straightforward approach could be using the noise of a content image as input and performing denoising with the pre-trained personalized diffusion model. While this method captures the target style, it often alters the original image's content. Previous work \cite{gu2024photoswap, gu2024swapanything} has shown that the self-attention feature maps in the early stages of denoising encode the image's content and structural information. Therefore, we first store the self-attention maps from the UNet during the content image reconstruction, treating them as content priors. During the generation of the target image, we replace the self-attention maps with these content priors. Since content and structural information are typically established in the early stages of denoising, this attention swapping operation is performed only within the first $k$ steps.

\paragraph{Implementation Details.}

We employ Stable Diffusion 1.4 \cite{rombach2022high} as our base model. We use BLIP-2 \cite{li2023blip} to generate the text prompt for the content image.
% , and the style prompt is in the form of ``a photo in the style of *". 
We apply random cropping and horizontal flipping to the input image to improve the robustness for style learning. Our model is trained on a single NVIDIA A6000 GPU with a batch size of 1 and a learning rate of 1e-6. 
The number of fine-tuning steps and time may vary slightly for each reference image, but on average, approximately 1500 steps are sufficient. We set $k$ = 25 for time-aware attention swapping. It takes about 10 seconds to generate a style-transferred image during the inference stage.

\section{Experiments}

% \begin{table*}[]

% \resizebox{\linewidth}{!}{
% \begin{tabular}{cccccccccc}
% \hline
% Methods & Ours & StyleID & InstantStyle & Diffuse-IT & InST & AesPA-Net & CAST & AdaAttN & ArtFlow \\ \hline
% % FID \downarrow   & \textbf{630}  & 470     & 221       & 213        & 89   & 322       & 583  & 379     & 414        \\ 
% LPIPS \downarrow   & 0.4923  & 0.5959     & 0.7226       & 0.7435        & 0.8199   & 0.5328       & 0.6485  & 0.4727     & 0.6922        \\ 
% User Study \uparrow   & \textbf{630}  & 470     & 221       & 213        & 89   & 322       & 583  & 379     & 414        \\ \hline
% \end{tabular}
% }
% \caption{Quantitative comparison with several state-of-the-art image style transfer methods.}
% \label{tab:user_study}
% \end{table*}

% \usepackage{graphicx}
\begin{table*}[]
\centering
\resizebox{0.9\textwidth}{!}{%
\begin{tabular}{cccccccccc}
\hline
Metrics\textbackslash{}Methods & Ours         & StyleID & InstantStyle & Diffuse-IT & InST & AesPA-Net & CAST           & AdaAttN & ArtFlow \\ \hline
Style Loss $\downarrow$   & \textbf{0.7641}  & 1.0902     & \underline{0.7653}       & 1.2312        & 0.9875   & 0.8766       & 0.7842  & 1.0288     & 0.7693        \\ 
LPIPS $\downarrow$   & \underline{0.5191}  & 0.5959     & 0.7226       & 0.7435        & 0.8199   & 0.5328       & 0.6485  & \textbf{0.4727}     & 0.6922        \\ 
\hline
Rank 1 (\%) $\uparrow$ & \textbf{9.1} & 5.06    & 2.76         & 2.85       & 1.01 & 2.20      & \underline{6.07}           & 3.49    & 3.49    \\
Top 3 (\%) $\uparrow$  & 14.54        & 13.61   & 6.16         & 6.25       & 2.48 & 10.40     & \textbf{17.75} & 13.71   & \underline{15.10}   \\ \hline
\end{tabular} 
}
\caption{Quantitative comparison with SOTA image style transfer methods.  The best results
are in \textbf{bold} while the second best results are marked with \underline{underline}.}
\label{tab:user_study}
\end{table*}

\subsection{Qualitative Comparison}
We compared multiple state-of-the-art methods to demonstrate the effectiveness of our approach. As shown in the third column of Figure \ref{fig:compare_other}, our method can well preserve and transfer the reference signature style. Specifically, our approach effectively transfers and preserves key signature style elements, such as the moon in the background in the first row and the small picture details in the second row of the style image, onto the content image.  
In contrast, other methods typically only transfer simple colors and fail to achieve high-quality  signature style transfer. Furthermore, for complex style images, our method generates more natural images compared to other methods, which often produce many artifacts (see the last four columns of the second row). Our method also maintains the structure of the content image well, whereas other tuning-based methods, such as InST \cite{zhang2023inversion} and DiffuseIT \cite{kwon2022diffusion}, alter the original content image's structure. This further demonstrates that our hypernetwork-powered style-aware tuning method can more precisely inverse style attributes, while the time-aware attention injection effectively preserves content information.

\subsection{Quantitative Comparison}
We randomly selected 10 content images and 15 style images, generating a total of 150 stylized images for each method by applying the Cartesian product to the content and style images. To evaluate content fidelity, we followed \cite{chung2024style} and employed the LPIPS metric \cite{zhang2018unreasonable}, which measures the similarity between the stylized image and the corresponding content image. \ye{ For style similarity, we adopted the Style Loss \cite{gatys2016image}, measuring the alignment between the stylized image and the corresponding style image.}

We compared our method with eight SOTA style transfer methods: StyleID, InstantStyle, Diffuse-IT, InST, AesPA-Net, CAST, AdaAttN, and ArtFlow. \ye{As presented in Table~\ref{tab:user_study}, our method achieves the second-lowest LPIPS score of 0.5191 and the lowest Style Loss of 0.7641, surpassing the majority of existing approaches.}
This result indicates that our approach effectively preserves content and style fidelity during signature style transfer. Furthermore, we also conducted a comprehensive user study. We randomly collected 12 content-style pairs, each producing nine images generated by the methods mentioned above. Participants were asked to select and rank the top three results from these nine randomly-arranged images based on two criteria: 1) the preservation of signature or key styles from the style image, including distinct and recognizable visual traits such as geometric patterns, color palettes, and brush strokes; 2) the preservation of content from the content image, including overall shape, structure, and semantics of the main object. A total of 1,152 votes were collected from 32 participants. As demonstrated in Table~\ref{tab:user_study}, we computed the proportion of first-place rankings (Rank 1) and the overall percentage of top three votes (Top 3) for each method. Our approach achieved the highest Rank 1 score (9.1\%) and a relatively high percentage of Top 3 votes (14.54\%). These results highlight that the style-transferred images generated by our method were preferred by users, showcasing its superior performance in signature style transfer.

\ye{As shown in Table \ref{tab:time}, we compare our method with tuning-based methods like InST and Diffuse-IT, which require 20 and 18 minutes for tuning, respectively. Our method is more efficient, with 10 minutes of tuning and 10 seconds of inference. Although tuning-free methods like StyleID, InstantStyle, and CAST are faster (5-8 seconds for inference), they do not preserve style details well. Our method achieves a better balance between speed and style quality.}

\begin{table}[t]
\centering
\resizebox{0.9\linewidth}{!}{
\setlength{\tabcolsep}{2pt}
\begin{tabular}{c c c | c c c}
\hline
Methods & Tuning & Inference & Methods & Tuning & Inference \\
\hline
InST & 20 mins & 24 s & StylelD & no need & 7 s \\
Diffuse-IT & 18 mins & - & InstantStyle & no need & 5 s \\
Ours & 10 mins & 10 s & CAST & no need & 8 s \\
\hline
\end{tabular}
}
\caption{\ye{Comparison of time consumption in the fine-tuning and inference phases between ours and other methods.}}
\label{tab:time}
\end{table}

\begin{figure}[t]
    \centering
    \includegraphics[width=0.7\linewidth]{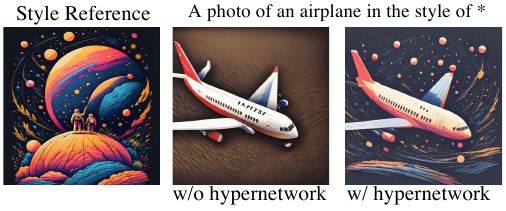}
    \caption{The ablation study on hypernetwork.}
    \label{fig:ablation_hypernetwork}
\end{figure}

\subsection{Ablation Study}

\paragraph{Hypernetwork.}
As shown in Figure \ref{fig:ablation_hypernetwork}, our hypernetwork effectively facilitates the precise learning and inversion of the style. In contrast, when fine-tuning without the hypernetwork, the generated images match the target text but fail to retain the reference style. This further demonstrates the effectiveness of our proposed hypernetwork.

\begin{figure}[t]
    \centering
    \includegraphics[width=0.87\linewidth]{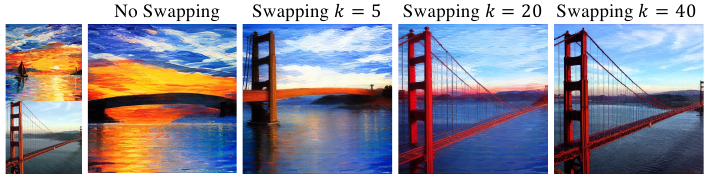}
    \caption{The ablation study on attention swapping.
    }
    \label{fig:ablation_attn_swap}
\end{figure}

\begin{figure}[t]
    \centering
    \includegraphics[width=0.87\linewidth]{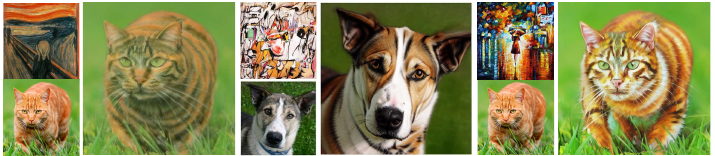}
    \caption{Local style transfer results of our method. The transfer area is the foreground animal region by default.}
 
    \label{fig:local_transfer}
\end{figure}
% \vspace{-10pt}
\begin{figure}[t]
    \centering
    \includegraphics[width=0.77\linewidth]{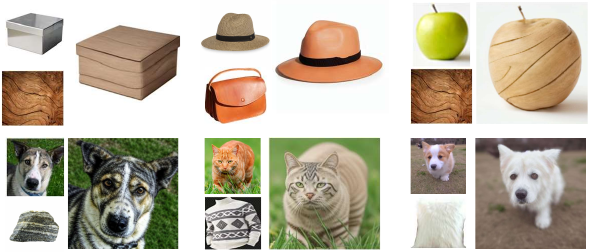}
    \caption{Cross-domain texture transfer of our method.
    }
    \label{fig:texture_transfer}
\end{figure}
% \vspace{-10pt}
\begin{figure}[!ht]
    \centering
    \includegraphics[width=0.77\linewidth]{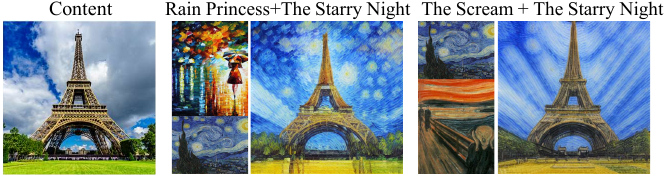}
    \caption{Style transfer based on multiple style references. 
    }
    \label{fig:2style_transfer}
\end{figure}

% \vspace{-15pt}
\begin{figure}[!ht]
    \centering
    \includegraphics[width=0.77\linewidth]{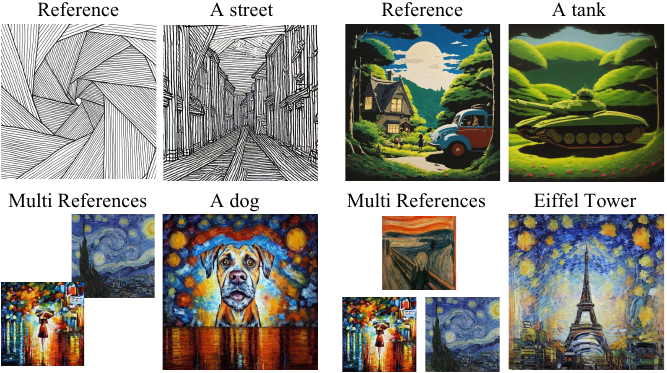}
    \caption{The style-guided text-to-image generation results.}
    \label{fig:style_guided}
\end{figure}

\paragraph{Attention Swapping.}
As shown in Figure \ref{fig:ablation_attn_swap}, when attention swapping is not used, the generated images only retain the target style but fail to maintain the image content. As the \(k\) value increases, the structure and other content-related information in the synthesized image become increasingly similar to the content image, while the style gradually diminishes. This demonstrates the effectiveness of the attention swapping technique in preserving content consistency.

% \vspace{-1.5cm}
% \paragraph{\textbf{Further Applications.}}
\subsection{Further Applications}
In addition to style transfer tasks, our method supports other applications including local style transfer, texture transfer, style fusion, and style-guided text-to-image generation.

% \paragraph{\textit{Local Style Transfer.}}
\paragraph{Local Style Transfer.}
Local style transfer applies style only to regions specified by a user-provided mask. Within the masked areas, we use SigStyle for style transfer, while denoising reconstruction is applied to non-masked areas to maintain consistency. Blending operations then integrate these regions seamlessly, producing a complete image and achieving local style transfer, as shown in Figure \ref{fig:local_transfer}.

% Local style transfer aims to apply style transfer only to the regions of interest specified by the user, while keeping other areas unchanged. This requires the user to provide a mask for the regions of interest in advance. Within the masked regions, we apply SigStyle for style transfer. For the areas outside the mask, we perform denoising reconstruction to ensure consistency. Finally, we perform blending operations to integrate these regions seamlessly, resulting in a complete image and achieving local style transfer, as shown in Figure \ref{fig:local_transfer}.

\paragraph{Texture Transfer.}
Texture, appearance, and style are interrelated concepts best learned by the same module, the UNet decoder. By replacing "style" with "appearance" in prompts while keeping inversion and transfer processes unchanged, a mask constrains the texture transfer region. Figure \ref{fig:texture_transfer} demonstrates high-quality cross-domain texture transfer, preserving the original image's pose, structure, identity, and other content.

% Texture, appearance and style are all visual abstract concepts with significant similarities. Therefore, these concepts should be learned by the same module, namely the UNet decoder. Based on this, we only need to modify the word "style" to "appearance" in the content and target text prompts, while keeping the inversion and transfer processes unchanged, and use a mask to constrain the texture transfer region. 
% Figure \ref{fig:texture_transfer} shows the results of high-quality cross-domain texture transfer, demonstrating our method can seamlessly transfer the reference texture onto the original image while preserving the original image's pose, structure, identity, and other content information.

\paragraph{Style Fusion.}
We can fuse multiple styles to create a new style for transfer and customized generation, leading to more intriguing results. This is achieved by fine-tuning with multiple style images. As shown in Figure \ref{fig:2style_transfer}, our method effectively transfers the fused style onto the content image. 

% For instance, we can generate an image with a sky that combines the styles of \textit{Rain Princess} and \textit{The Starry Night}, or create a scene featuring the sky from \textit{The Starry Night} with the Eiffel Tower in the style of \textit{The Scream}.

\paragraph{Style-guided Text-to-Image Generation.}
Our fine-tuning mechanism represents style as a special token *, enabling style-guided text-to-image generation. With a single style image, we can generate images guided by that style (see the first row of Figure \ref{fig:style_guided}). When using multiple style images, our method fuses them into a new style for more creative outputs (see the second row of Figure \ref{fig:style_guided}).

\section{Conclusion}In this paper, we presented SigStyle, a novel framework for high-quality signature style transfer using only a single style reference image. By introducing a hypernetwork-powered style-aware fine-tuning mechanism, our approach enhances the accuracy and efficiency of style inversion while addressing severe single-image overfitting issues. Additionally, our time-aware attention swapping technique ensures content consistency during the style transfer process. Extensive experiments show that SigStyle outperforms existing methods in preserving signature styles and supports various applications, including global and local style transfer, texture transfer, and style-guided text-to-image generation etc. 
% SigStyle encounters efficiency challenges, necessitating a distinct tuning process for each style.
% While achieving promising results, our SigStyle still has some limitations.
Currently, SigStyle still needs to fine-tune the diffusion model for each given style image during inference, which limits its deployment on resource-constrained devices.
How to further reduce the computation cost of the style learning and transferring process will be worthy to investigate.
Moreover, it will also be interesting to explore how to use more specified text prompt to guide more refined and more controllable style transfer.

\section*{Acknowledgments}
This work is supported in part by the National Natural Science Foundation of China (No. 62202199, No. 62406134) and the Science and Technology Development Plan of Jilin Province (No. 20230101071JC).

\bigskip

\bibliography{aaai25}

\end{document}